\documentstyle [prl,aps,preprint]{revtex}
\begin{document}
\draft
\title{Quantum state sensitivity to initial conditions}
\author{Gonzalo Garcia de Polavieja\footnote{Electronic mail: 
gonzalo@rydberg.thchem.ox.ac.uk}} 
\address{Physical and 
Theoretical Chemistry Laboratoty, South Parks Road, Oxford OX1 3QZ, U.K.}
\maketitle
\begin{abstract}
The different time-dependent distances of two arbitrarily close quantum or
classical-statistical states to a third fixed state are shown to imply an
experimentally relevant notion of state sensitivity to initial conditions. 
A quantitative classification scheme of quantum states by their sensitivity 
and instability in state space is given that reduces to the one performed by 
classical-mechanical Lyapunov exponents in the classical limit.
\end{abstract} 
Stimulated by the pioneering work of Peres \cite{Peres84,PeresBook}
on state sensitivity to small changes in the Hamiltonian, several new
developments have taken the study of quantum state sensitivity to the
experimental realm. Schack and Caves 
\cite{Schack92} have analyzed the sensitivity 
to perturbations as the amount of information about a perturbing environment
that is needed to keep entropy from increasing. Their approach links the
study of quantum sensitivity to statistical mechanics and to quantum
information theory. Recently, Gardiner \textit{et al. }\cite{Gardiner} have
proposed a detailed experimental set-up of an ion trap to study quantum
state sensitivity to changes in the Hamiltonian. A different approach has
been taken by Ballentine and Zibin \cite{Ballentine} that study the
emergence of classical state sensitivity from quantum theory in
computational time reversal. The difficulty of directly tackling the problem
of quantum state sensitivity to initial conditions stems from the following
reason. Classical sensitivity to initial conditions is made precise when
measured by Lyapunov characteristic exponents \cite{Lasota} 
\begin{equation}
\lambda_{c}({\bf x}_{0})=\lim_{d_{c}(0)\rightarrow 0}
\lim_{
t\rightarrow \infty 
}\frac{1}{t}\ln
\left( \frac{d_{c}\left( {\bf x}_{t},{\bf y}_{t}\right) }{d_{c}\left( 
{\bf x}_{0},{\bf y}_{0}\right) }\right)   \label{1}
\end{equation}
with $d_{c}\left( {\bf x}_{t},{\bf y}_{t}\right) $ the Euclidean
distance \cite{explanation} between two phase space vectors ${\bf x}%
_{t}$ and ${\bf y}_{t}$. There is classical sensitivity to initial
conditions when $\lambda _{c}({\bf x}_{0})>0$ and the trajectory starting
at ${\bf x}_{0}$ is said to be unstable. A direct attempt to define the
quantum analog of the classical Lyapunov characteristic exponent in (\ref{1}%
) substitutes the phase space distance between two classical trajectories by
the Hilbert-space $\cal{H}$ distance between two close quantum vectors
giving 
\begin{equation}
\lambda_{\mathcal{H}}\left[ \left| \psi (0)\right\rangle \right]
=\lim_{d_{\mathcal{H}}(0)\rightarrow 0}
\lim_{t\rightarrow \infty }
\frac{1}{t}\ln \left( \frac{d_{\mathcal{H}%
}\left( \left| \psi (t)\right\rangle ,\left| \varphi (t)\right\rangle
\right) }{d_{\mathcal{H}}\left( \left| \psi (0)\right\rangle ,\left| \varphi
(0)\right\rangle \right) }\right) =0,  \label{2}
\end{equation}
a zero quantum Lyapunov exponent for all states as any two states do not
separate at all with $d_{\mathcal{H}}\left( \left| \psi (t)\right\rangle
,\left| \varphi (t)\right\rangle \right) \equiv \left\| \left| \psi
(t)\right\rangle -\left| \varphi (t)\right\rangle \right\| $ $=\left\|
\left| \psi (0)\right\rangle -\left| \varphi (0)\right\rangle \right\| $ and 
$\left\| \cdot \right\| =\sqrt{\left\langle \cdot \right. \left| \cdot %
\right\rangle }$. According to (\ref{2}) all possible quantum states are
stable and no sensitivity of quantum states to initial conditions can be
found.

 However, there are three strong objections against the use of $%
\lambda _{\mathcal{H}}$ as a measure of state sensitivity to initial
conditions and instability. Firstly, the quantum mechanical state space is
the complex projective Hilbert space $C{\cal P}^{n}$ that has a curvature
not present in the complex Hilbert space ${\mathcal H}_{C}^{n+1}.$ It is a
nontrivial K\"{a}hler manifold with a symplectic form and an associated
Riemannian metric \cite{Marsden}. Secondly, equation (\ref{2}) is obtained
by substituting the classical phase space trajectory distance in (\ref{1})
by a quantum distance. This is not a mere technical error but a conceptual one
as the closest classical object to a quantum state is a Liouville density,
not a single trajectory 
\cite{PeresBook,Schack92,Ballentine}. 
Recall for example that in
the classical limit the dynamical equation for the Wigner function or for the
coherent state representation of the quantum density reduce to the Liouville
equation \cite{PeresBook}. A definition analogous to (\ref{2}) for
Liouville densities gives $\lambda _{\mathcal{L}}\left[ \rho(0)\right]=0$
with $d_{\mathcal{L}}\left( \rho _{1}(t),\rho _{2}(t)\right) \equiv \sqrt{%
\int dqdp\rho _{1}^{1/2}(q,p,t)\rho _{2}^{1/2}(q,p,t)}=d_{\mathcal{L}}\left(
\rho _{1}(0),\rho _{2}(0)\right) $ by Koopman$^{\prime }$s theorem. The fact
that $\lambda _{\mathcal{L}}(\rho (t))=0$ for all possible Liouville
densities tells us that we cannot know from (\ref{2}) if quantum states have
sensitivity as this type of measure for classical Liouville densities is
insensitive to classical-mechanical instability. Thirdly, unlike the
classical trajectory distance, $d_{\mathcal{H}}$ is a bounded metric, thus
precluding the exponential divergence. In this paper we propose to overcome
the above three objections in three steps to provide a working scheme to
classify quantum and classical-statistical states by their sensitivity and instability
in projective space ${\cal P}^{n}.$ This sensitivity and instability are
shown to be measurable by the different transition probabilities of two
nearby states to a third fixed state.

\textit{Step 1}. A classical density $\rho (x)$ can be mapped into a real
Hilbert space ${\cal H}_{R}^{n+1}$ by taking its square root, $\psi
_{c}(x) \equiv \sqrt{\rho (x)}.$ The expectation of a classical operator $F$ is $%
\left\langle F\right\rangle =\left\langle F\psi _{c}^{2}\right\rangle
/\left\langle \psi _{c}^{2}\right\rangle ,$ which means that the physical
space is not ${\cal H}_{R}^{n+1}$but the space of equivalence classes
obtained by identifying $\psi _{c}\sim \lambda \psi _{c}$ for any $\lambda
\in {\bf R}-\left\{ 0\right\} $. This physical space is the real
projective space ${\bf R}{\cal P}^{n}.$ Similarly, in quantum
mechanics the true physical space is not the complex Hilbert space ${\cal 
H}_{C}^{n+1}$ but the space of equivalent classes obtained by identification
of vectors $\psi _{Q}\sim \lambda \psi _{Q}$ for any $\lambda \in {\bf C}%
-\left\{ 0\right\} ,$ the complex projective space ${\bf C}{\cal P}%
^{n}.$ Classical and quantum states in ${\cal P}^{n}$ will be written as $%
\widetilde{\psi }$. The distance between two points in ${\cal P}^{n}$ (real or complex) can be defined as
\begin{equation}
d_{\mathcal{P}}\left[ \widetilde{\psi }_{1,}\widetilde{\psi }_{2}\right]
=2\arccos \left| \left\langle \frac{\psi _{1}}{\left\| \psi _{1}\right\| },%
\frac{\psi _{2}}{\left\| \psi _{2}\right\| }\right\rangle \right| ,
\label{metric}
\end{equation}
which is the length of the geodesic connecting the two points in ${\cal P}^{n}$ as measured by the Fubini-Study metric \cite{Marsden,Anandan}.

\textit{Step2}. Now we are faced with the problem of finding relevant
distances to characterize the motion of classical-statistical and quantum
systems by their sensitivity and stability properties. According to the
quantum expectation rule \cite{PeresBook}, the probability that a quantum
state prepared in state $\widetilde{\psi }_{S}$ will pass succesfully the
test for the state $\widetilde{\psi }_{R}$ is $\left| \left\langle 
\widetilde{\psi }_{S}\right. \left| \widetilde{\psi }_{R}\right\rangle
\right| ^{2}.$ This shows that the distances of  two arbitrarily close
states $\widetilde{\psi }_{S}$ and $\widetilde{\psi }_{S}+\delta \widetilde{%
\varphi }$ to a third fixed state $\widetilde{\psi }_{R}$ are experimentally
relevant as from (\ref{metric}) $\left| \left\langle \widetilde{\psi }%
_{S}\right. \left| \widetilde{\psi }_{R}\right\rangle \right| ^{2}=\cos
^{2}\left( d_{\mathcal{P}}\left[ \widetilde{\psi }_{S,}\widetilde{\psi }%
_{R}\right] /2\right) $ and similarly for $\left| \widetilde{\psi }%
_{S}+\delta \widetilde{\varphi }\right\rangle $ . These distances $d_{%
\mathcal{P}}\left[ \widetilde{\psi }_{S}(t),\widetilde{\psi }_{R}\right] $
and $d_{\mathcal{P}}\left[ \widetilde{\psi }_{S}(t)+\delta \widetilde{%
\varphi }(t),\widetilde{\psi }_{R}\right] $ are different and time-dependent
and their difference is expected to behave differently in stable and
unstable systems.

\textit{Step 3}. The distance between two states 
in (\ref{metric}) is bounded, $d_{\mathcal{P}}
\in \left[ 0,\pi \right] ,$ which obviously
precludes exponential divergence. To understand how to study unstable
systems in terms of a bounded metric we express the unbounded phase space
Euclidean distance, $d_{c}\in \left[
0,\infty \right) ,$ in terms of a bounded distance, say $d_{c}^{b}
\in \left[ 0,\pi \right] ,$ that preserves the
topology ($d_{c}^{b}$ is $0$ iff $d_{c}$ is $0$) as \cite{Giles} 
\begin{equation}
d_{c}^{b}\left( {\bf x},{\bf y}\right) \equiv \pi d_{c}\left( 
{\bf x},{\bf y}\right) /\left(1+d_{c}\left( {\bf x},{\bf y}\right)\right) 
\label{bounded}
\end{equation}
The classical-mechanical Lyapunov exponent in (\ref{1}) can then be
expressed in terms of the classical bounded metric in (\ref{bounded}) as

\begin{equation}
\lambda_{c}({\bf x}_{0})=
\lim_{\Lambda_{c}(0)\rightarrow 0}
\lim_{t\rightarrow \infty }\frac{1}{t}\ln
\left( \Lambda _{c}\left( {\bf x}_{t},{\bf y}_{t}\right) /\Lambda
_{c}\left( {\bf x}_{0},{\bf y}_{0}\right) \right) .  \label{Lyapunov2}
\end{equation}
with $\Lambda _{c}\left( {\bf x},{\bf y}\right) $ the \textit{%
divergence} function that is the unbounded phase space distance expressed in
terms of the bounded one as 
\begin{equation}
\Lambda _{c}\left( {\bf x},{\bf y}\right) \equiv 
d_{c}^{b}\left( {\bf x},{\bf y}\right)/ \left(\pi -d_{c}^{b}\left( 
{\bf x},{\bf y}\right)\right) .
\end{equation}

We are now ready to overcome the three objections to the definition in (\ref
{2}) and propose an analogue of the classical-mechanical Lyapunov exponent
in ${\cal P}^{n}$ (that we name $\mathcal{P}$-Lyapunov exponent). Let $%
\widetilde{\psi }_{S}(t)$ and $\widetilde{\psi }_{S}(t)+\delta \widetilde{%
\varphi }(t)$ be two arbitrarily close system states and $\widetilde{\psi }%
_{R}$ a reference state on the path $\widetilde{\psi }_{S}(t),$ $\widetilde{%
\psi }_{R}=\widetilde{\psi }_{S}(\tau )$ $.$ The $\mathcal{P}$-Lyapunov
exponent is obtained by the different distances of the two arbitrarily close
system states to the reference point of the form 
\begin{equation}
\lambda _{\mathcal{P}}^{\infty }\left[ \widetilde{\psi }_{S}(0)\right]
=\lim_{\delta\rightarrow 0}
\lim_{t\rightarrow \infty }\frac{1}{t}\ln \left| \frac{\Lambda \left[ 
\widetilde{\psi }_{S}(t)\right] -\Lambda \left[ \widetilde{\psi }%
_{S}(t)+\delta \widetilde{\varphi }(t)\right] }{\Lambda \left[ \widetilde{%
\psi }_{S}(0)\right] -\Lambda \left[ \widetilde{\psi }_{S}(0)+\delta 
\widetilde{\varphi }(0)\right] }\right|  \label{LyapunovQ}
\end{equation}
with $\Lambda \left[ \widetilde{\psi }_{S}(t)\right] $ the \textit{divergence%
} function for the distance between the system state $\widetilde{\psi }%
_{S}(t)$ and the reference state $\widetilde{\psi }_{R}$%
\begin{equation}
\Lambda \left[ \widetilde{\psi }_{S}(t)\right] \equiv d_{\mathcal{P}%
}\left[ \widetilde{\psi }_{R},\widetilde{\psi }_{S}(t)\right] /\left(\pi -d_{%
\mathcal{P}}\left[ \widetilde{\psi }_{R},\widetilde{\psi }_{S}(t)\right]\right) 
\label{divergence}
\end{equation}
and $d_{\mathcal{P}}$ the bounded distance between states in (\ref{metric}).
There is sensitivity to the initial condition $\widetilde{\psi }_{S}(0)$ of
degree $\lambda _{\mathcal{P}}^{\infty }
$ when $\lambda _{\mathcal{P}}^{\infty }
>0$ and the path $\widetilde{\psi }_{S}(t)$ is said to
be unstable. The calculation of $\lambda _{\mathcal{P}}^{\infty }
$ can be obtained 
with a single path by choosing $\widetilde{%
\psi }_{S}(t)+\delta \widetilde{\varphi }(t)=\widetilde{\psi }_{S}(t+\delta t).$

We have obtained the definition of $\mathcal{P}$-Lyapunov exponent in (\ref
{LyapunovQ}) by overcoming the three objections to the definition in (\ref{2}%
). We now put this definition to the test. The approach taken to find the $%
\mathcal{P}$-Lyapunov exponent has been kinematic and it is expected to be
valid for classical-statistical and quantum states. First we show its
validity for classical-statistical states and its equivalence to the
classical-mechanical Lyapunov exponent for several examples. Consider the classical unstable map 
$\Phi (x)=rx$ with $r>1$,  
for which the classical-mechanical Lyapunov characteristic exponent is $%
\lambda _{c}({\bf x}_{0})=\ln r.$ After $n$ steps the classical classical
density evolves to $\rho _{n}(x)=P^{n}\rho (x)=\rho \left( x/r^{n}\right) /r^{n}$.
Take as initial system state a square-looking function $\widetilde{\psi }%
_{S}(x,0)=\sqrt{\rho _{0}(x)}=\sqrt{\Theta (x)-\Theta (x-b)}$
and $\Theta $ the Heaviside function . The distance between the system state and the reference
state $\widetilde{\psi }_{R}=$ $\widetilde{\psi }_{S}(0)$ is found to be 
$d _{\mathcal{P}}=2 \arccos | r^{-n/2}| $. Its corresponding \textit{divergence} function (\ref
{divergence}) ultimately diverges exponentially and 
the $\mathcal{P-}$Lyapunov exponent in (\ref{LyapunovQ}) can be calculated
to be 
\begin{equation}
\lambda _{\mathcal{P}}^{\infty }\left[ \widetilde{\psi }_{S}(0)\right] =
\ln(r)/2=\lambda _{c}({\bf x}_{0})/2,
\end{equation}
half the classical-mechanical Lyapunov exponent. There is then sensitivity to initial conditions
for classical statistical states and they are classified as unstable in the
same manner than using classical-mechanical Lyapunov exponents for all $r$. Figure 1(a)
shows the convergence of $\lambda _{\mathcal{P}}^{t}$ to $\lambda _{\mathcal{%
P}}^{\infty }$ for different values of $r$.

A more sophisticated map is the $r$-adic transformation 
$\Phi (x)=rx \, (mod 1)$ with $r>1,$ which is an \textit{exact} transformation \cite{Lasota}
(therefore \textit{mixing}) with classical-mechanical Lyapunov exponent $%
\lambda _{c}({\bf x}_{0})=\ln r$. A necessary feature added to
instability for \textit{mixing} systems is boundedness that implies, as in
the classical-mechanical trajectory case, that the distance between the system state and the reference state is
 saturated at the value $2\arccos| \left( \left\langle \psi _{S}\right\rangle \left\langle \psi
_{R}\right\rangle \right)|$
, but before this saturation takes place the $\mathcal{P}$-Lyapunov exponent
can be extracted to be arbitrarily close to $\ln (r)/2=
\lambda _{c}({\bf x}_{0})/2$ by choosing a sufficienty localized initial 
classical-statistical state. A classical
transformation related to the $r$-adic map of interest for us in the later
discussion of quantum states is the baker's map, a map of the unit square
onto itself of the form 
\begin{equation}
\Phi (x,y)=(2x-[2x],(y+[2y])/2)  \label{baker}
\end{equation}
where $[\cdot ]$ indicates integer part. This map is area-preserving and 
\textit{mixing} with classical-mechanical Lyapunov exponent $\ln 2.$
Following the same steps as in the $r$-adic map, we find the same result
with $r=2$.

We have obtained classical-statistical state sensitivity to initial
conditions and instability for classical-mechanical unstable and \textit{%
mixing} maps. Encouraged by these results we now proceed to study
quantum systems. To test our definition of $\mathcal{P-}$Lyapunov exponent
for quantum states we first choose simple analytic cases, the parabolic
barrier and the harmonic oscillator. All states of the harmonic oscillator
(see Figure 1(b) for a Gaussian function) are stable, $\lambda _{\mathcal{P}%
}^{\infty } =0.$ For the parabolic
barrier, with potential $V(x)=-m\omega ^{2}x^{2}/2,$ and
classical-mechanical Lyapunov exponent $\lambda _{c}({\bf x}_{0})=\omega $
there are stable and unstable quantum states. The most stable ones are the
eigenstates $\phi _{\alpha }$ and in fact any discrete sum of eigenstates $%
\psi _{S}=\sum_{n}c_{n}\phi _{n}$ is stable. On the other hand, states that
use the continuous spectrum are unstable. For example, the distance between a Gaussian initial
state on top of the barrier, $\widetilde{\psi }_{S}(x)=\exp \left( -m\omega
_{0}x^{2}/2\hbar \right) ,$  and the reference state is of the form $2 \arccos | \left( \cosh
^{2}\omega t+\left( \omega /\omega _{0}-\omega_{0} /\omega\right) ^{2}\sinh
^{2}\omega t\right) ^{-1/4}|$. Its corresponding \textit{divergence}
function (\ref{divergence}) ultimately diverges exponentially with time and 
the $\mathcal{P}$-Lyapunov
exponent in (\ref{LyapunovQ}) is found to be 
\begin{equation}
\lambda _{\mathcal{P}}^{\infty }\left[ \widetilde{\psi }_{S}(0)\right]
=\omega /2=\lambda _{c}({\bf x}_{0})/2.
\end{equation}
Therefore, in contrast to the starting definition (\ref{2}), there is
quantum state sensitivity to initial conditions and quantum unstable paths.
Figure 1(b) shows the convergence of $\lambda _{\mathcal{P}}^{t}$ to the $%
\mathcal{P}$-Lyapunov exponent $\lambda _{\mathcal{P}}^{\infty }$ for two
states with different values of $\omega $.

We now consider the added complication of\textit{\ }classically \textit{%
mixing} systems for which we will use the classical baker map in (\ref{baker}%
). Quantization of a classical map is not unique and we consider two quantum
baker maps to illustrate two different types of behaviour. First we
understand the map as defined in configuration space, as did Mendes \cite{Mendes} for the
Arnol'd cat map, for which a quantum theory can be defined by the unitary Floquet
operator $U$ such that $\widetilde{\psi }_{S}(x,y)\rightarrow \widetilde{\psi }_{S}^{\prime }(x,y)=
\widetilde{\psi }_{S}\left( \Phi ^{-1}(x,y)\right) =U\widetilde{\psi }
_{S}(x,y)$
with $\Phi $ in (\ref{baker}) and $\widehat{p}%
_{x,y}=-i\hbar \nabla _{x,y}$ the momentum. For this quantum map the
situation is analogous to that of classical-statistical states under the
baker map and one then finds state sensitivity and unstable quantum paths in
the same way. The second quantization procedure that we consider understands
the classical baker map in position-momentum space and we adopt the
Balazs-Voros-Saraceno \cite{Balazs,Saraceno} quantization procedure.
The new characteristic that interests us is the discreteness of the
spectrum. As we have already found for the parabolic barrier, it is possible
to have stable states made of a discrete sum of eigenstates 
\cite{Bocchieri,Hogg,Brown,Radons} even when localized onto classically
unstable structures \cite{Garcia}. However, this does not pose a problem
for the correspondence principle as for systems that appear classical to us
the recurrence times of the quantum states are longer than the age of the
Universe \cite{Peres82}. In typical numerical and experimental results the
recurrence times due to the discreteness of the spectrum might be longer
than that due to classical boundedness by the energy hypersurface. We relax
the mathematical convenience of considering asymptotic
definitions and consider coverged values for experimentally relevant times. Observe in Figure 1
that nonasymptotic values already give a classification of states by their
sensitivity and instability for finite times.
We now proceed to study the Balazs-Voros-Saraceno quantum map. Position and
momentum are discretized $q_{j}=p_{j}=\left( j+1/2\right) /N$ for $%
j=0,...,N-1$ with $N$ even. The transformation between the position basis $%
\left\{ \left| q_{j}\right\rangle \right\} $ and the momentum basis $\left\{
\left| p_{j}\right\rangle \right\} $ is given by $\left( G_{N}\right) _{ij}=\left\langle p_{i}\right| \left.
q_{j}\right\rangle$ 
and the baker map is then defined as 
\begin{equation}
\widetilde{B}\equiv G_{N}^{-1}\left( 
\begin{array}{ll}
G_{N/2} & 0 \\ 
0 & G_{N/2}
\end{array}
\right) ,
\end{equation}
where the matrices $G_{N/2}$ map position kets on the lower (upper) half of
the position interval to the lower (upper) part of the momentum interval and 
$G_{N}^{-1}$ maps back to the position representation. We choose as initial
quantum state $\widetilde{\psi }_{S}\left( q_{j}\right) =\exp \left( -\left(
q_{0}-q_{j}\right) ^{2}/2\alpha +ip_{0}q_{j}/\alpha \right) $ with $\alpha
=10^{4}$ and $q_{0}=p_{0}=0.003$. Figure 2(a) shows the finite-time $%
\mathcal{P}$ -Lyapunov exponent for $N=1800$ and times shorter than the
bounded time $t_{b}\approx 8$ and it is found to be close to half the
classical-mechanical Lyapunov exponent $\ln (2)/2$ as in the previous
examples.

The notion of $\mathcal{P}$ -Lyapunov
exponent introduced in this paper can be measured in many types of
experimental set-ups. It can be measured in optical filtering experiments by
the different reduction of intensity of two nearby states. In molecular
experiments the distance between the system state and the reference state
can be obtained from the Fourier transform of the Franck-Condon spectrum.
Very recently, Gardiner \textit{et al}. \cite{Gardiner} have proposed an
ion trap set-up as a practical realization of a delta-kicked harmonic
oscillator to measure the Peres \cite{Peres84} overlap $O(t)=\left|
\left\langle \psi _{S}\right| \widehat{U}_{1}^{\dag }(t)\widehat{U}_{2}(t)
\left| \psi _{S}\right\rangle \right|^{2}$ with $\widehat{U}_{1}$
and $\widehat{U}_{2}$ two close unitary evolution operators to study the
state sensitivity to changes in the Hamiltonian. We can reinterpret their
results as sensitivity to initial conditions in the following manner.
Consider the two nearby vectors $\left| \psi _{S}(t)\right\rangle =\widehat{U%
}_{1}^{-1}(t)\widehat{U}_{2}(t)\left| \psi _{S}(0)\right\rangle $ and $%
\left| \psi _{S}(t+dt)\right\rangle .$ Their distances from the
reference vector $\left| \psi _{R}\right\rangle =\left| \psi
_{S}(0)\right\rangle $ are given in terms of the Peres overlap by $2 \arccos O(t)$. Figure
2(b) shows the \textit{divergence} function for the two states of Fig.4(b)
and (d) of \cite{Gardiner}, respectively. Figure 2(b) also shows the $%
\mathcal{P}$ -Lyapunov exponent and we find that one path is unstable with $%
\lambda _{\mathcal{P}}\approx 0.017$ and the other stable with $\lambda _{%
\mathcal{P}}\approx 0.$

The classification performed by the quantum $%
\mathcal{P}$-Lyapunov exponent will be in general different from the
classical-mechanical one, but they must coincide in the classical limit. To
see this first note that the Wigner representation or coherent state
representation of the quantum state reduce to the classical Liouville
density in the classical limit
\cite{PeresBook}. Moreover, the classification performed by
the classical-statistical (Liouville) $\mathcal{P}$-Lyapunov exponent coincides with the
one given by classical-mechanical Lyapunov exponents. To see this note that
the distance between two classical densities given by (\ref{metric}) when
the underlying trajectories obey ${\bf x}(t)={\bf x}(0)\exp (\lambda _{c}t)$, with $%
\lambda _{c}$ the classical-mechanical Lyapunov exponent, is obtained to be $%
d_{\mathcal{P}}\left[ \widetilde{\psi }_{S},\widetilde{\psi }_{R}\right]
=\arccos \left( \exp \left( -\lambda _{c}t\right) \right) $ and the classical
Liouville $\mathcal{P}$-Lyapunov exponent reduces to $\lambda _{\mathcal{P}%
}\left[ \widetilde{\psi }_{S}(0)\right] =\lambda _{c}/2$ 
. This property opens the possibility of studying in
detail the classicalization of instability, most interestingly by the
calculation of the quantum $\mathcal{P}$-Lyapunov exponent when adding
classical or quantum noise and with increasing dimensionality of the
system.

To conclude, a successful measure of quantum and classical-statistical state
sensitivity to initial conditions and instability has been obtained. The
classification given for quantum states reduces in the classical limit to
that performed by classical-mechanical Lyapunov exponents. This
classification has been studied in several examples, including recently proposed ion trap
states.

J.I. Cirac, M.S. Child, S. Gardiner, R. Schack, E. Sj\"{o}qvist and P. Zoller are acknowledged for useful discussions. A Marie Curie fellowship is also acknowledged.

\newpage

\section*{Figure Captions}

\textbf{Figure 1}. Classification of states by their sensitivity to initial
conditions as measured by the $\mathcal{P}$ -Lyapunov exponent in (\ref
{LyapunovQ}) for (a)\ classical-statistical states under the unstable map $\Phi (x)=rx$
 with $r=2,3$ and $5$ converging to half the
classical-mechanical Lyapunov exponent  $\ln (r)/2$ and (b) for quantum
states in the harmonic oscillator and the parabolic barrier with $\omega =2$
and $5$ converging to half the classical Lyapunov exponent $0,1$ and $2.5$,
respectively. 

\textbf{Figure 2}. Classification of quantum states with discrete spectrum
by their finite-time $\mathcal{P}$ -Lyapunov exponent. Pannel (a) shows  the
finite-time $\mathcal{P}$ -Lyapunov exponent for the quantum baker map
close to $\ln (2)/2.$ The probability density is shown as an inset.
Pannel (b) shows the \textit{divergence} function (\ref{divergence}) for a
state centered on a classically regular region and a state centered on a
classically chaotic region of a delta-kicked harmonic oscillator, indicated by (1) and (2) respectively. The inset
shows their corresponding finite-time $\mathcal{P}$ -Lyapunov exponent that
converges approximately to 0 and 0.017. The data for (b) has been obtained
from the experimental proposal of Gardiner \textit{et al}.\cite{Gardiner}
on an ion trap set-up.

\end{document}